\documentclass[preprint]{vgtc}               

\ifpdf
  \pdfoutput=1\relax                   
  \usepackage{graphicx}                
  \DeclareGraphicsExtensions{.pdf,.png,.jpg,.jpeg} 
\else
  \usepackage{graphicx}                
  \DeclareGraphicsExtensions{.eps}     
\fi%

\graphicspath{{figures/}{pictures/}{images/}{./}} 

\usepackage{microtype}                 
\PassOptionsToPackage{warn}{textcomp}  
\usepackage{textcomp}                  
\usepackage{mathptmx}                  
\usepackage{times}                     
\usepackage{cite}                      
\usepackage{tabu}                      
\usepackage{booktabs}                  

\onlineid{0}

\vgtccategory{Research}

\vgtcinsertpkg

\preprinttext{To appear in IEEE VIS 2022 Workshop on Visualization for Social Good on October 16, 2022 in Oklahoma City, OK, USA.}

\title{Representing Marginalized Populations: Challenges in Anthropographics}

\author{Priya Dhawka\thanks{e-mail: priyadarshinee.dhawk@ucalgary.ca} %
\and Helen Ai He\thanks{e-mail:helen.he1@ucalgary.ca}
\and Wesley Willett\thanks{e-mail:wesley.willett@ucalgary.ca}} %
\affiliation{
  \scriptsize{University of Calgary}
  }

\abstract{Anthropographics are human-shaped visualizations that have primarily been used within visualization research and data journalism to show humanitarian and demographic data. However, anthropographics have typically been produced by a small group of designers, researchers, and journalists, and most use homogeneous representations of marginalized populations---representations that might have problematic implications for how viewers perceive the people they represent. In this paper, we use a critical lens to examine  anthropographic visualization practices in projects about marginalized populations. We present critiques that identify three potential challenges related to the use of anthropographics and highlight possible unintended consequences---namely (1) creating homogeneous depictions of marginalized populations, (2) treating marginalization as an inclusion criteria, and (3) insufficiently contextualizing datasets about marginalization. Finally, we highlight opportunities for anthropographics research, including the need to develop techniques for representing demographic differences between marginalized populations and for studies exploring other potential effects of anthropographics.} 

\CCScatlist{
  \CCScat{Human-centered computing}{Visu\-al\-iza\-tion}{Visualization design and evaluation methods}{}
}

\begin{document}


\firstsection{Introduction}

\maketitle

Demographic data includes information about personal characteristics such as age, gender, and race, as well as a wide range of other factors. This type of data is often closely tied to experiences of marginalization, as the collection and use of demographic data involves grouping individuals based on their  characteristics---approaches which have long been (and are still) used to discriminate and marginalize people through systems of oppression such as ageism, racism, and sexism. Demographic characteristics are now also considered to be legally protected attributes in countries such as Canada and the USA, which has implications for how this data is used ~\cite{OFCCP,DOJCanada}. Here, we use the term \textbf{marginalized populations} to refer to groups that have been excluded from participating in economic, educational, social, political, and other aspects of mainstream life, usually on the basis of their demographic characteristics. Additionally, given systems of oppression, working with marginalized populations in visualization research usually involves unique considerations from designers and researchers.

One interesting new avenue for visualizing these kinds of data about people data is  \textbf{anthropographics}---visualizations that use human-shaped marks to represent people~\cite{Boy, Morais2020, Morais2021}. This research has focused primarily on using anthropographics to represent data about marginalized populations (especially data about human rights issues) with the goal of evoking empathy towards them. Anthropographics have also appeared in a growing number of data journalism stories to show data about people---including representing COVID-19 deaths in the USA~\cite{NYTcovid2020}, Afghan evacuees being resettled in US states~\cite{AP2021}, and mass shooting statistics in the USA~\cite{WP2018}. 

Yet, the use of anthropographics presents a number of potential challenges.
For example, despite showing data about diverse populations, anthropographic projects tend to use simple human shapes that encode little demographic information. Where they show any variation between individuals, it is usually limited to a few narrow gender (male/female) and age categories (child/adult), potentially creating homogeneous or dehumanizing representations of marginalized populations~\cite{Boy, Morais2021, NYTcovid2020, WP2018}. In doing so, these graphics can ignore the social, political, and historical contexts surrounding the treatment of marginalized populations, treating groups as generic and interchangeable.
Moreover, the visualization community's enthusiasm for using anthropographics to provoke empathy may lead designers to use marginalization as an inclusion criteria, resulting in over-representation of projects that show marginalized populations as the objects of suffering.
Furthermore, a small group of researchers, designers, and data journalists are involved in the creation of these representations---few of whom hail from these marginalized groups~\cite{WEIRDCHI}. As a result, this small community likely lacks the background necessary to contextualize data about many of the diverse populations being represented.

In this paper, we describe three potential issues that may arise when representing marginalized populations via anthropographics. We also reflect on some potential effects of these practices and discuss the complexities of working with data about marginalization. Finally, we highlight several possible opportunities for future  anthropographics research. 

\section{Background and Related Work}
Within information visualization, relatively little canonical guidance exists for designers visualizing data about human populations. However, recent work like Schwabish and Feng's Do No Harm guide for data practitioners~\cite{SchwabishFeng}---which was primarily derived from interviews with visualization creators in the US---advocates for inclusive visualization design practices  and an awareness of racial equity. For example, Schwabish and Feng recommend considering privacy when data for marginalized populations is below a certain threshold, and examining the language or colors used to communicate racial or gender categories. More broadly, Dörk and colleagues propose considering visualization approaches that prioritize contingency, disclosure, empowerment, and plurality when evaluating visualizations~\cite{dork}. Dörk et al. strongly emphasize that visualizations are situated, are influenced by the intentions of the designer, and also frequently make the invisible visible. Within data science, D'Ignazio and Klein proposed a data feminism framework to help address concerns around power imbalances when working with data and marginalized populations~\cite{datafem}. Similarly, Gebru et al. have proposed \textit{Datasheets for Datasets} to help researchers working in machine learning address concerns around potential data bias. Datasheets for Datasets include guidelines for researchers to consult with domain experts when datasets, such as data about marginalized populations, require additional consideration~\cite{gebru}.

Recent thinking on visualization ethics and audiences also suggests that unintended effects of visualization design choices can strongly influence how viewers relate to visualizations. For instance, Correll mentions the phenomenon of viewers experiencing feelings of alienation from the people being represented by a visualization~\cite{Correll}. Similarly, from their study of data receptivity among participants in rural Pennsylvania, Peck et al. concluded that for some viewers, familiarity with data is an important but extremely personal factor that influences how they relate to a visualization~\cite{peck}. 

Anthropographics represent a distinct point in the space of visualizations about people, and visualization research has mostly sought to assess the degree to which these human-shaped visualizations elicit certain prosocial emotions. Early work by Boy and colleagues proposed a design space for anthropographics that involved simple human shaped icons~\cite{Boy}. They also conducted experiments to gauge the effectiveness of anthropographics at eliciting empathy by measuring how participants allocated donations to humanitarian causes. This study used simple human shapes to encode demographic data about children living through the Syrian crisis and experiencing crises ranging from poverty, internal displacement, and access to water and education~\cite{Boy}. Subsequent work by Morais et al. has proposed an alternative design space for anthropographics that includes seven dimensions, focusing on what information the anthropographic shapes communicate (granularity, specificity, coverage, authenticity) and how they show it (realism, physicality, situatedness)~\cite{Morais2020}. Morais et al. also conducted a number of experiments with anthropographics and offered perspectives from a study measuring anthropographics' ability to evoke empathy towards marginalized migrant populations. This study involved data about migrant deaths, with participants primarily from Europe being asked to allocate money to humanitarian causes to measure the effectiveness of anthropographics in eliciting prosocial feelings~\cite{Morais2021}. Morais et al. also used the same simple human shapes to represent migrants from two different geographical regions and who died of different causes. 
Similarly, in their experimental immersive anthropographic visualization titled ``A Walk among the Data'', Ivanov et al. encoded the gender and age of victims of mass shootings in the US but did not include identifiable physical characteristics due to lack of data about the populations being represented~\cite{ivanov2019walk}. Currently, work in anthropographics has not yielded conclusive results on whether or not such visualizations can be used to evoke prosocial feelings.

\section{Critiques}
Based on examinations of a number of recent anthropographics from both visualization research and data journalism, we outline three critiques of current practices in anthropographics. We also identify some potential effects of these practices on the representation of marginalized populations.

\subsection{Homogeneous Representations of Marginalization}
Anthropographic visualizations usually do not include identifiable details of the individuals being represented. Rather, most examples use a small set of human-shaped templates which are repeated to represent larger groups of people. The use of similar human-shaped icons and visuals to represent distinct marginalized populations erases the demographic differences between these groups and treats marginalized populations as  interchangeable. Although as McCloud argues ~\cite{mccloud}, homogeneous human shapes are abstractions that can be understood as representing people generally,  the use of generic human shapes to represent marginalized populations with different demographic characteristics can obscure distinctly different cultures and lived experiences. Furthermore, representing marginalized populations as interchangeable can influence how they are publicly perceived, which might affect the implementation of policies directed towards them ~\cite{clark2020disproportionate,kantamneni2020impact}. 

For instance, in Figure 1, the choice to represent migrants from Southeast Asia and migrants from the Middle East using the same shapes~\cite{Morais2021} can confound the demographic and cultural differences between these two groups. Representing these populations as interchangeable hides the unique experiences of marginalization, causes of death, and humanitarian issues each face. 
Similarly, in Figure 2, the use of similar visuals to represent many demographically-distinct victims of the COVID-19 pandemic  erases important differences such as social class, race, and gender~\cite{NYTcovid2020}---despite the fact that many of these demographic differences also likely influenced their COVID exposure and cause of death ~\cite{clark2020disproportionate,kantamneni2020impact}.

\begin{figure}[t]
 \centering 
 \includegraphics[width=\columnwidth]{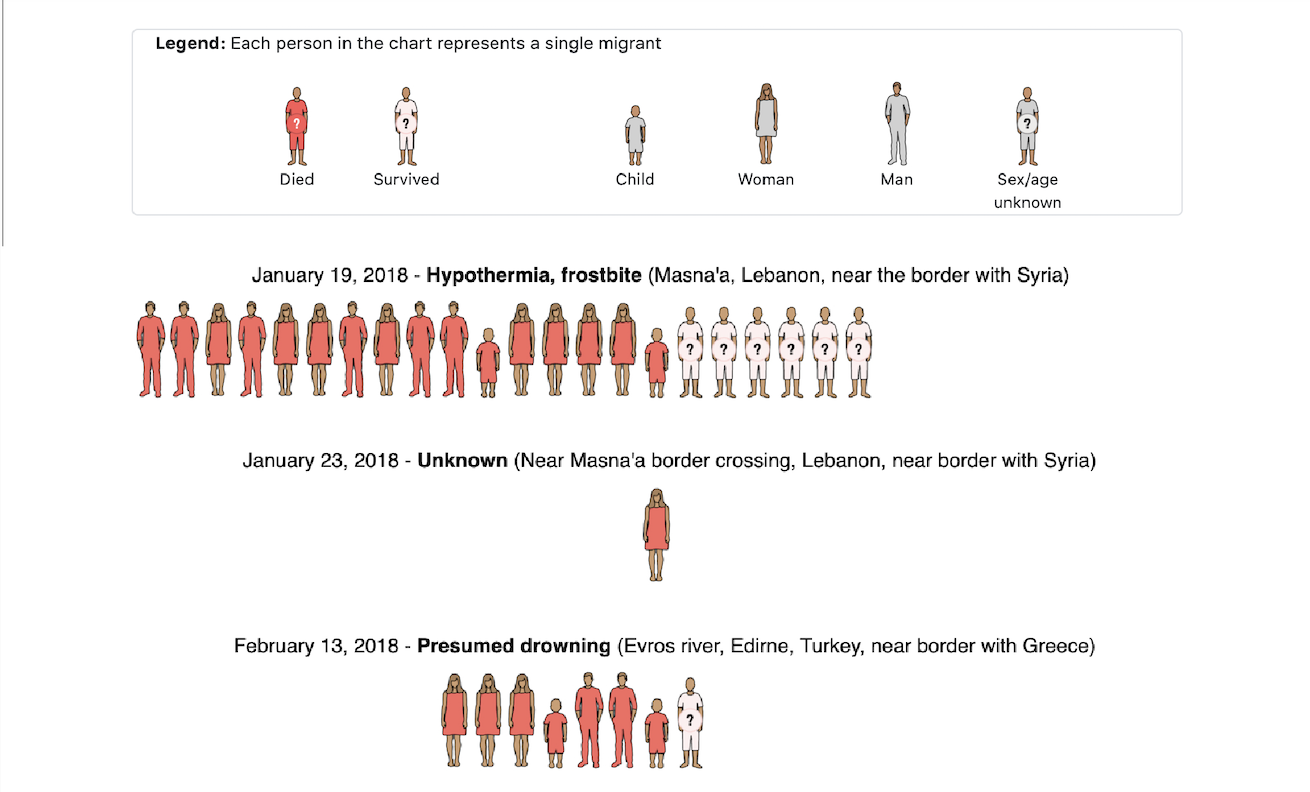}
 \caption{``Can Anthropographics Promote Prosociality? A Review and Large-Sample Study'' by Morais et al\cite{Morais2021}. A set of human-shaped icons is used to represent migrants from the Middle East. Each icon depicts a person. These visualizations are also used to represent migrants from Southeast Asia in the same study. }
\end{figure}

\begin{figure}[t]
 \centering 
 \includegraphics[width=\columnwidth]{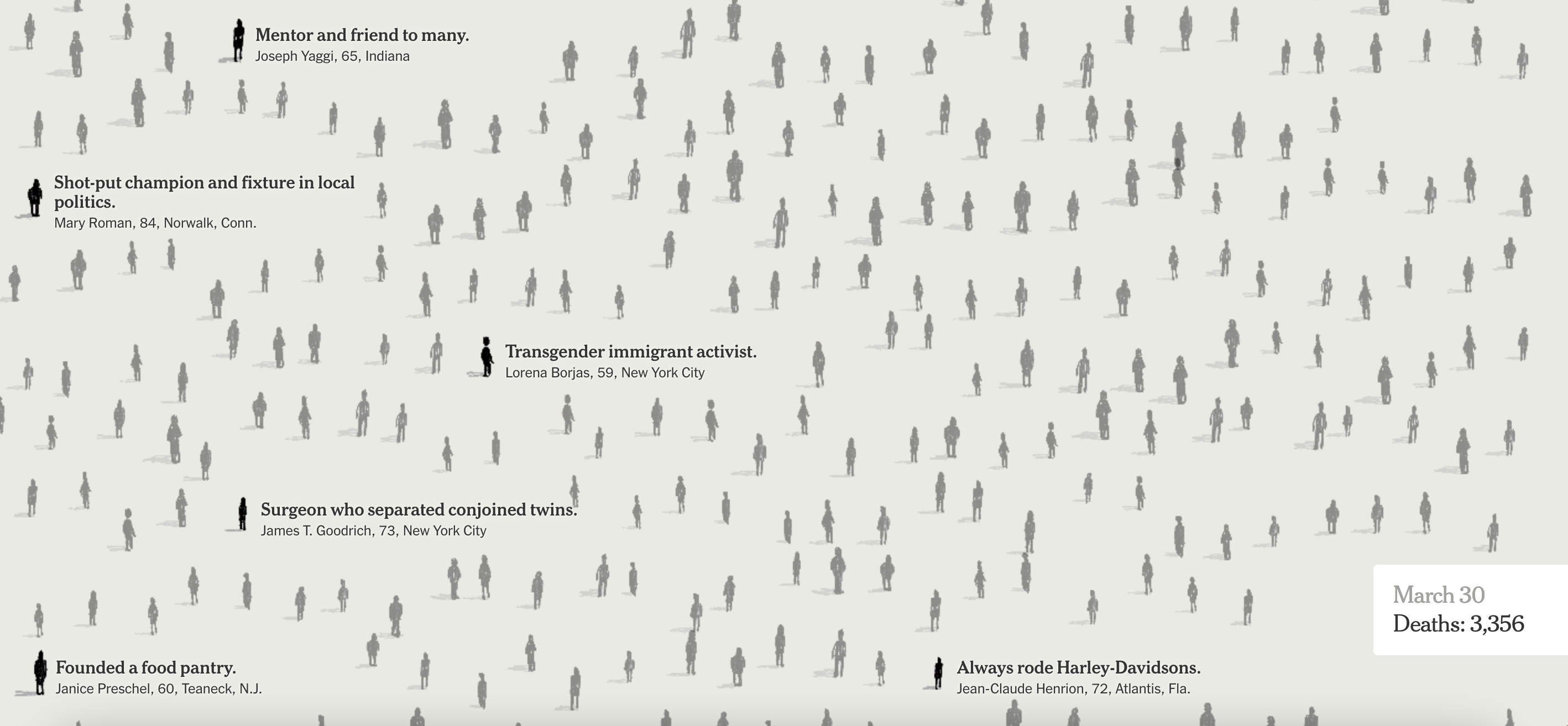}
 \caption{``Remembering the 100,000 lives lost to coronavirus in America'', published in the New York Times in 2020\cite{NYTcovid2020}. Human silhouettes are used to represent individuals who died from COVID-19 in the US. Each silhouette represents a person. Data and individual details were collected from US newspaper obituaries.}
\end{figure}

\subsection{Marginalization as Inclusion Criteria}
The availability of data about populations can also influence which anthropographic projects they are included in. 
Because data about marginalized populations may be incomplete or require special considerations for privacy and anonymity---demographic data and other contextualizing information about these groups is often sparse. 
Meanwhile, the data that \textit{is} available about marginalized populations (including groups like refugees, unhoused people, or abuse victims) often focuses exclusively on the issues affecting them---such as conflicts, death, and suffering. This can result in visualizations that repeatedly and exclusively focus on marginalized groups through the lens of victimhood.
For instance, data stories about past and ongoing conflicts around the world have sought to use anthropographics to raise awareness about the state of marginalized populations. Yet, designers of these data stories have often used graphic images of death and starvation alongside anthropographics, charts, or maps---showing these populations in demeaning positions and reinforcing stereotypes about their marginalization~\cite{Guardian2022}. One recent example is an article by The Guardian that reports on famine resulting from current conflicts in countries such as Ethiopia, Somalia, South Sudan, Yemen and Afghanistan that includes graphic images of starvation juxtaposed with charts showing data about affected populations~\cite{Guardian2022}.

Issues surrounding marginalization as an inclusion criteria are not unique to visualization research. In fact, there has been recent interest in journalism in re-examining how persistent depictions of discrimination and suffering (instead of positive stories about educational, social, and political achievements) can normalize dehumanizing narratives of marginalization~\cite{Time2022}.

\subsection{Contextualizing Datasets about Marginalization}
The relative lack of visualization designers and researchers from marginalized groups also influences how data about marginalization is understood within complex social, historical, and political contexts~\cite{WEIRDCHI}. 
In particular, designers without sufficient cultural context may risk reinforcing stereotypes or caricatures of marginalized populations. 
One example is a data physicalization project titled ``Slave Voyages: Reflections on Data Sculptures'' that shows the journey of enslaved people from Africa to the Americas (Figure 3). In this project, the researchers used colored beads to represent data about enslaved people---including whether or not they survived the transatlantic journey~\cite{kominsky}.
However, the researchers were unaware that glass beads were used to barter for enslaved people and were significant for their Black viewers until after they completed the project~\cite{kominsky}. Because of the lack of cultural background and domain knowledge, there was little consideration for the significance of the chosen material and how viewers might perceive the design choices in the context of historical and familial connections to slavery. 

Similarly, the data journalism article from the Associated Press seen in Figure 4 uses a color palette ranging from gray to dark brown to show the distribution of Afghan refugee resettlment across the US~\cite{AP2021}. This color palette had been used extensively in other data stories not involving marginalized people, but in this specific context the graphic's beiges and browns can easily be misinterpreted as skin tones. This choice of color palette is problematic considering how non-White immigrants are often stereotypically perceived in the US. Specifically, darker tones might reinforce racist stereotypes about immigrants changing the racial demographic composition of a state to be less White. Here, an awareness of racial equity, as Schwabish and Feng emphasize~\cite{SchwabishFeng}, would have been useful to consider how the choice of color palette might be received in the context of the broader history of immigration to the US.

\begin{figure}[bt]
 \centering 
 \includegraphics[width=\columnwidth]{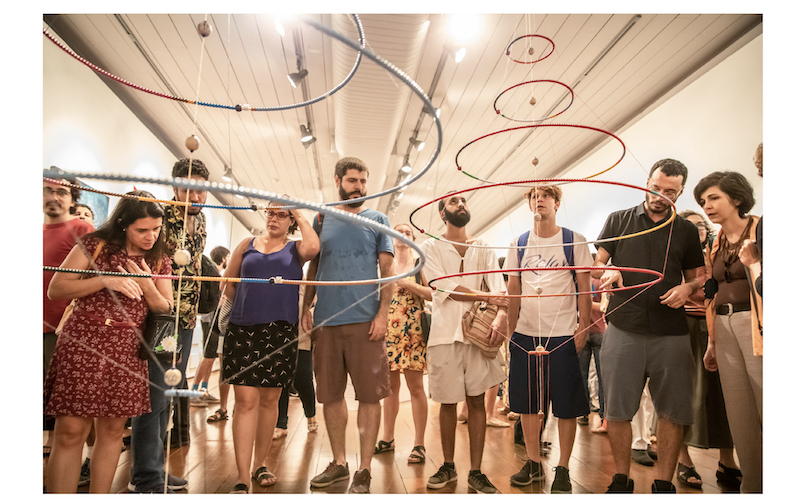}
 \caption{``Slave Voyages: Reflections on Data Sculptures'' by Kominsky and de Oliveira\cite{kominsky}. This data sculpture uses colored beads to represent enslaved people, the locations in Africa where they were captured, and the locations they were taken to in the Americas.}
\end{figure}
\begin{figure}[tb]
 \centering 
  \includegraphics[width=\columnwidth]{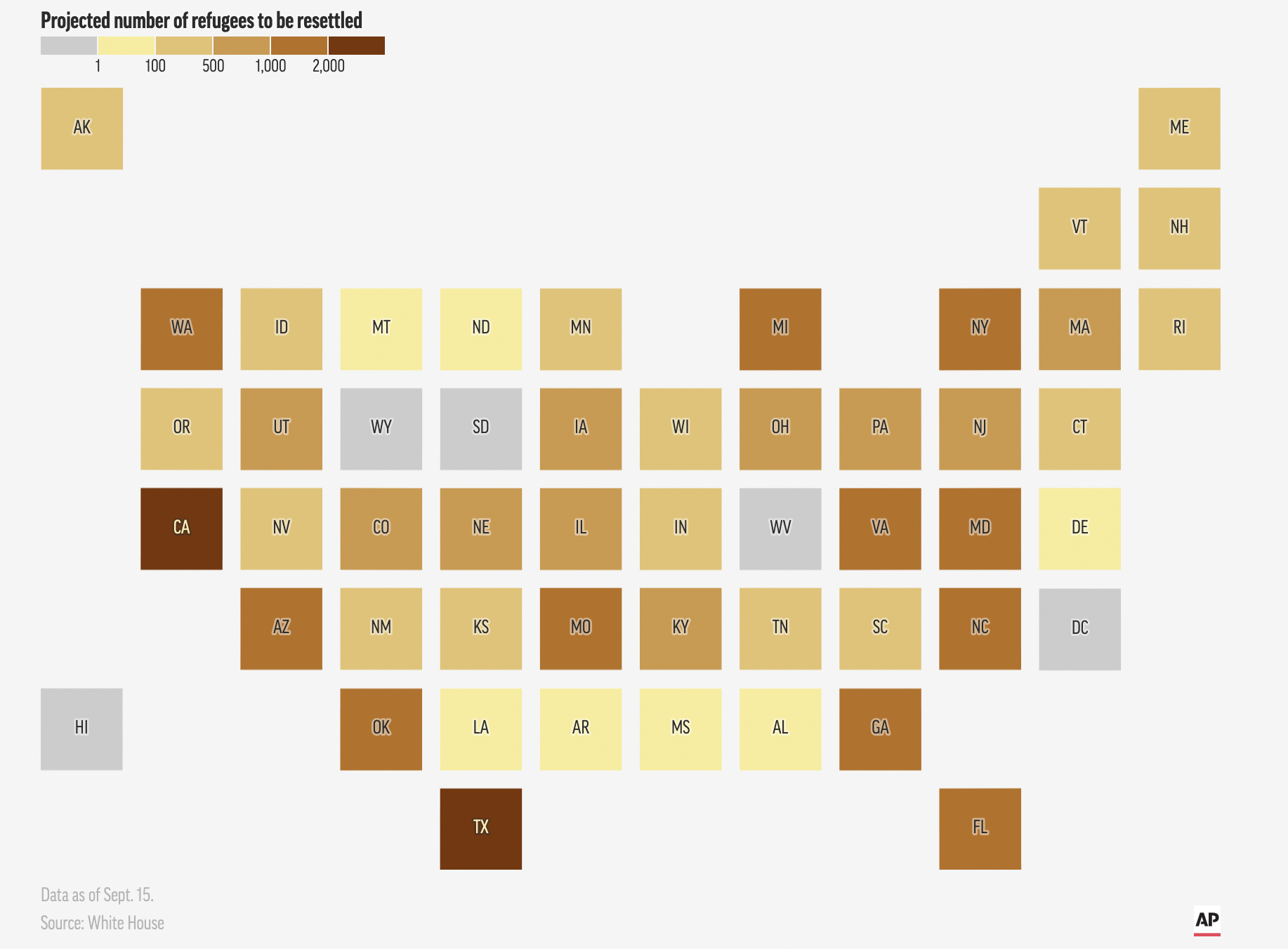}
 \caption{``States learning how many Afghan evacuees coming their way'', published by The Associated Press in 2021\cite{AP2021}. The number of Afghan refugees to be resettled in each US state is represented by varying shades of brown.}
\end{figure}

\section{Opportunities for Anthropographics Research}
In light of these challenges, we highlight two potential future research directions for anthropographics.

\subsection{Representing Differences between Populations}
Most current anthropographics do not convey the demographic differences between marginalized groups. Yet, representing these demographic differences positively can help address issues surrounding homogeneous anthropographics. The process of creating representative, contextualized, and humanizing anthropographics of marginalized populations would almost certainly benefit from input from the populations being represented. We thus advocate for approaches such as participatory design that include marginalized populations in the design process of anthropographics. More broadly, designers and researchers should also seek to understand how marginalized populations perceive homogeneous anthropographics and how (or if) they wish to be represented.

While non-homogeneous anthropographics have the potential to provide more equitable representations of marginalized populations, we also acknowledge concerns around the practicality and usefulness of such representations. Often  researchers and designers do not have access to demographic data or good visual mappings for encoding it. Even where demographic data is present, creating humanizing and representative anthropographics can be challenging. If executed poorly, such approaches can even result in misrepresentation, harmful stereotypes, and caricatures. Creating distinct anthropographics for every demographic group being represented can also be time consuming for larger population-scale datasets such as the dataset about COVID-19 deaths in the USA (Figure 2). Despite these challenges, we emphasize that the creation of non-homogeneous anthropographics is a worthwhile endeavor to equitably represent the diversity among marginalized populations.

\subsection{Exploring Other Effects of Anthropographics}
Anthropographics have largely been used in projects that have attempted to provoke prosocial feelings towards marginalized populations. However, there are other potential uses of anthropographics that have yet to be explored---including supporting better understanding of complex demographic datasets, helping people to see themselves or people like them in the data, and creating opportunities for storytelling and engagement. Anthropographics have also not been evaluated for unintended side effects. For instance, homogeneous anthropographics could possibly alienate viewers from the people being represented or further harmful stereotypes about marginalized populations being interchangeable. More positive effects of anthropographics could include ensuring equitable representation of marginalized populations and dispelling stereotypes.

That said, the use of representative, non-homogeneous anthropographics may be limited to projects that involve adequate demographic data. These projects require considerations for the privacy of individuals being represented, particularly when anthropographics are realistic. Additionally, the types of data used might impact where anthropographics might be most useful, how these visualizations are evaluated, and what side effects are observed. On the other hand, demographic data is used in sectors such as financial services, healthcare, education, housing, immigration, and video games, among many others~\cite{Andrus}. This opens up new avenues for exploring creative ways of designing non-homogeneous anthropographics for diverse uses.

\section{Conclusion}

How we choose to communicate data about marginalized people to broader audiences can have implications for how these populations are perceived. As visualization researchers and designers, the visualizations we create are powerful, and even more so when these visualizations involve marginalized populations. Because of this, the concerns described in this paper are important for the visualization community to consider---especially given the risk of inadvertently harming the populations we seek to humanize. These issues stemming from global systems of oppression also permeate our research. An understanding of marginalization therefore can be useful to question the impact of visualization practices in projects for social good. Our goal with this work is to convey that marginalized populations are more than their experiences of marginalization, and we can create space for inclusion within visualization research.

\bibliographystyle{abbrv-doi}

\bibliography{references}
\end{document}